\documentclass[twocolumn,prc,floatfix,showpacs,preprintnumbers,amsmath,amssymb,merge,superscriptaddress]{revtex4-1}

\usepackage{graphicx}
\usepackage{dcolumn}
\usepackage{bm}
\usepackage{natbib}

\usepackage[utf8]{inputenc}

\begin{document}

\title{Cluster Radioactivity in Super Heavy Nuclei}
\author{M. Warda}
\email{michal.warda@umcs.pl}
\affiliation{Katedra Fizyki Teoretycznej, Uniwersytet Marii Curie--Sk\l odowskiej, Lublin, Poland}
\author{A. Zdeb}
\affiliation{Katedra Fizyki Teoretycznej, Uniwersytet Marii Curie--Sk\l odowskiej, Lublin, Poland}
\affiliation{Departamento de F\'\i sica Te\'orica, Universidad Aut\'onoma de Madrid, Spain}
\author{L. M. Robledo}
\affiliation{Departamento de F\'\i sica Te\'orica, Universidad Aut\'onoma de Madrid, Spain}

\date{\today}

\begin{abstract}
Cluster radioactivity is an exotic nuclear decay observed in actinides 
where a light nucleus is emitted while the remaining heavy mass residue 
is the doubly magic $^{208}$Pb or a nucleus in its neighborhood. We have 
investigated this type of decay in heavier nuclei up to Lv $(Z=116)$ 
within a microscopic theory. It has been found that super asymmetric 
fission with $^{208}$Pb as heavy fragment may be dominant decay channel 
in some super heavy nuclei. This reaction is closely related with 
cluster radioactivity.
\end{abstract}


\maketitle




{\it Introduction.}
In a seminal experiment carried out by Rose and Jones back in 1984 
\cite{Rose1984245} a new type of nuclear decay was discovered. Immersed 
in an enormous $\alpha$ decay background produced by the mother nucleus 
$^{223}$Ra a few events producing $^{14}$C were observed. The 
phenomenon received the name of cluster radioactivity (CR) due to the 
intermediate mass of the light fragment emitted. In the following 
years, 20 other cluster emitters have been discovered, see reviews in 
Refs. \cite{Bonetti1999643,Poenaru2002543081}. In this type of decay 
light nuclei ranging from $^{12}$C to $^{34}$Si are emitted by light 
actinides from $^{221}$Fr to $^{242}$Cm. The remaining heavy mass 
fragment in all these reactions is either the doubly magic $^{208}$Pb 
or one of its neighbours in the chart of nuclides. For this reason 
the phenomenon is also called lead radioactivity. The dominant double 
magic structure of the heavy fragment clearly shows the strong 
influence of shell effects on CR. As compared to other decay channels, 
CR is an exotic process: typical branching ratios to the dominant 
$\alpha$ emission are as low as $10^{-6} -10^{-12}$ and consequently 
the half-lives range from $10^{15}$ s to $10^{25}$ s. For nuclei 
heavier than $^{242}$Cm spontaneous fission becomes a competing 
channel increasing the difficulty to detect CR products among the 
numerous fission fragments present in the background. As a consequence, 
no CR have been observed in nuclei around mass $A=250$.

CR is usually described in the spirit of the Gamow model of $\alpha$ 
decay. In this model it is assumed that a preformed cluster made of 
several nucleons tunnels through a barrier created by the nuclear and 
Coulomb potentials 
\cite{Zdeb2013,Tavares1402-4896-86-1-015201,poenaruPhysRevC.83.014601,poenaru0954-3899-39-1-015105}. 
The exponential dependence of the tunneling probability with the 
parameters of the barrier leads to a modified Geiger-Nutall law 
\cite{renprc,qi,niPhysRevC.78.044310,BalasubramaniamPhysRevC.70.017301} 
relating half-lives for CR to the $Q$ value of the reaction. This 
approach presents two main disadvantages: first, it requires a model to 
estimate the preformation probability of the cluster and second, a 
local fit of additional model parameters is required to reproduce 
observed decay half-lives. An alternative to this model is to treat CR 
as a very asymmetric fission process to be described with the tools of 
the traditional fission decay model. This was the approach followed in 
Ref. \cite{Sandulescu1980528} to predict the existence of CR a few 
years before its experimental discovery. The most important ingredient 
of any fission model is the analysis of the changes in energy of the 
nucleus as it changes deformation in its way to scission 
\cite{krappe2012,Schunck2016}. In the context of CR, it was shown 
\cite{War11} that a super asymmetric fission valley can be found on the 
potential energy surface (PES) spanned by the quadrupole and octupole 
moments. It leads to a scission point with $^{208}$Pb or a nucleus of a 
very similar mass as one of the fission fragments. In the light 
actinides the fission barrier associated to this channel reaches a 
height of 25 MeV which is the right order of magnitude to reproduce CR 
half-lives.

Several authors have already suggested that SH elements may decay 
through CR 
\cite{PoenaruPhysRevLett.107.062503,poenaruPhysRevC.85.034615,zhangPhysRevC.97.014318, 
santosh}. Calculations performed within phenomenological approaches and 
semi-empirical formulae show a trend to predict shorter half-lives for 
CR in the SH region. Therefore, the competition with other 
disintegration channels becomes relevant and might have an impact on 
the very limits of the periodic table \cite{Nazarewicz2018} or in the r-process
nucleo-synthesis \cite{MARTINEZPINEDO2007199}.

The aim of this work is to study the possible existence of CR in SH 
nuclei within a fully microscopic theory. To this end we have used the 
selfconsistent Hartree-Fock-Bogoliubov model along with the Gogny D1S 
interaction to calculate the PES and the collective inertias required 
for the evaluation of CR half-lives. As relevant collective coordinates 
the axially symmetric quadrupole and octupole moments are used. 
This is a well-established quantum mechanics approach that allows to 
describe a very rich variety of nuclear shapes not limited by the 
number of deformation parameters. This type of calculations has been successfully 
applied both for the description of CR \citep{Robledo2008204} and 
fission in heavy \cite{War02} and SH nuclei 
\cite{wardaPhysRevC.86.014322, BARAN2015442}.

One of the common features of fission and CR is that the $Z$ and $N$ 
values of their fragments approximately conserve the $N/Z$ ratio of 
the mother nucleus. Therefore, to investigate the possibility to 
observe CR in heavier nuclei, one has to focus on possible emitters 
with the $N/Z$ value of $126/82=1.537$ corresponding to the always 
present heavy fragment of $^{208}$Pb. Therefore, we have chosen for our 
studies a set of even-even isotopes, one for each element, with an 
$N/Z$ ratio close to $1.537$. In this way we have selected 
$^{224}$Ra, $^{228}$Th, $^{234}$U, and $^{238}$Pu, where CR has already 
been observed. Heavier isotopes include $^{244}$Cm, $^{248}$Cf, 
$^{254}$Fm, $^{258}$No, $^{264}$Rf, $^{268}$Sg, $^{274}$Hs, $^{278}$Ds, 
$^{284}$Cn, $^{290}$Fl, and $^{294}$Lv. The final part of this chain 
belongs to the region of super heavy (SH) elements close to the 
isotopes experimentally produced in hot fusion reaction 
\cite{oganessian0954-3899-34-4-R01}. It is worth mentioning that 
$^{284}$Cn is the heaviest isotope that has been observed decaying 
through fission both at GSI and in Dubna 
\cite{DuellmannPhysRevLett.104.252701,oganessian2011synthesis}. These 
are difficult experiments due to the low production rate and a total 
number of only 28 fission events have been reported for this isotope.


\begin{figure}
\includegraphics[width=1\columnwidth, angle=0]{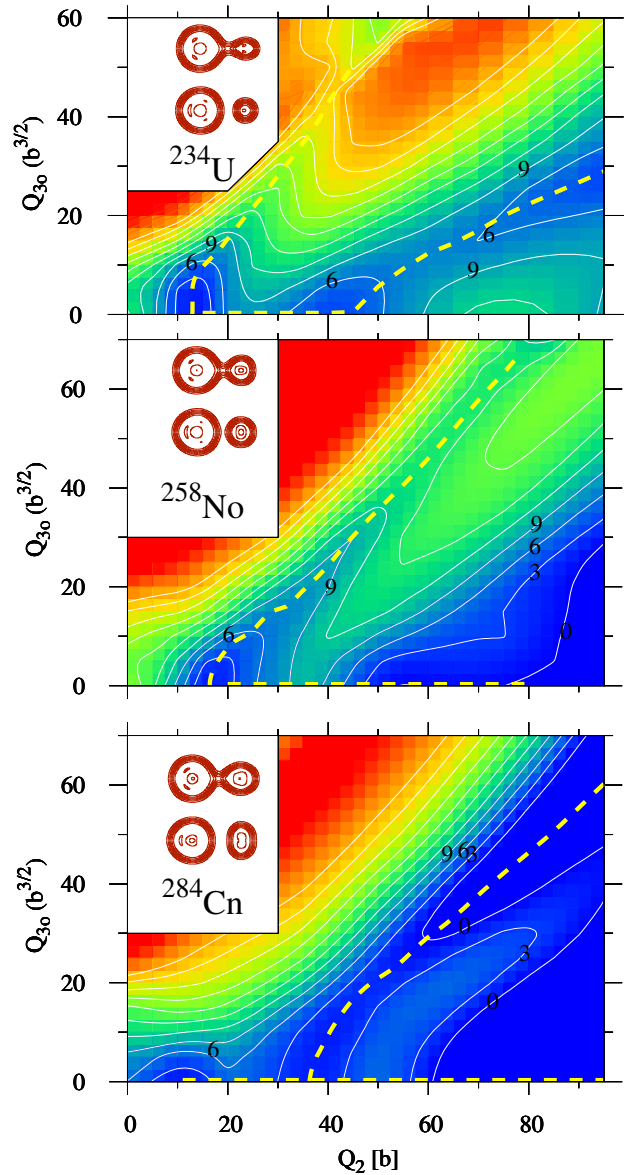}
\caption{PES of (a) $^{234}$U, (b) $^{258}$No, and (c) $^{284}$Cn. Constant energy lines are plotted every 3 MeV. Fission paths are marked with yellow dashed lines. Insets show pre- and post-scission configurations. \label{PES}}
\end{figure}
 
 \begin{figure}
\includegraphics[width=0.9\columnwidth, angle=0]{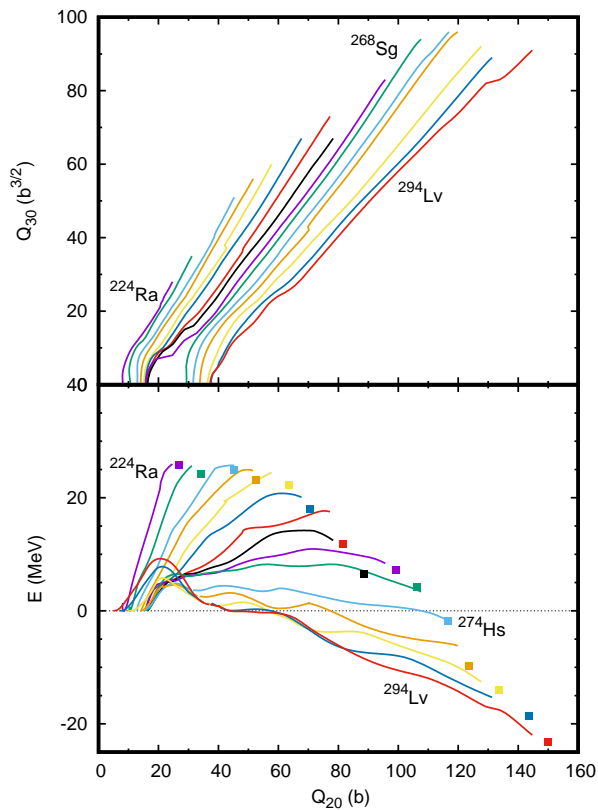}
\caption{ (a) CR fission paths in quadrupole and octupole coordinates for isotopes: $^{224}$Ra, $^{228}$Th, $^{234}$U, $^{238}$Pu, $^{244}$Cm, $^{248}$Cf, $^{254}$Fm, $^{258}$No, $^{264}$Rf, $^{268}$Sg, $^{274}$Hs, $^{278}$Ds, $^{284}$Cn, $^{290}$Fl, and $^{294}$Lv. (b) Fission barriers of aforementioned nuclei. Squares correspond to scission configuration calculated with Eq. (\ref{EC}) with $Q$ values obtained from experimental masses \cite{Wang1674-1137-36-12-003}, wherever possible, and liquid drop systematics \cite{MOLLER20161} \label{PATHS}}
\end{figure}

{\it Results.}
A typical example of a nucleus where 
CR has been observed is the light actinide $^{234}$U. In the PES of this 
nucleus, shown in Fig. \ref{PES}a, the asymmetric fission 
path, typical for heavy nuclei up 
to Fm is easily found. It goes from the ground state through the first 
symmetric fission barrier, which is 6.6 MeV high, to reach the fission isomer. 
Then it crosses
the second asymmetric saddle and gets to the scission point. On 
the same PES another valley can be noticed. It begins at the ground 
state and heads directly towards octupole deformed shapes. Along most 
of the whole fission path in this valley both quadrupole and octupole 
moments increase simultaneously up to $Q_{20}=45$ b and $Q_{30}=51 
\;\mathrm{b}^{3/2}$. In the saddle point the energy reaches 26.6 MeV. 
The scission point is located at the saddle point. From there on the 
PES corresponds to the Coulomb energy of the two fragments as they 
drift away and the energy decreases hyperbolically with the 
increasing distance between the fragments. The nuclear matter density 
distributions before and after scission are plotted on the inset of 
Fig. \ref{PES}a. From them it is easy to find that mass and shape of 
the heavy fragment corresponds to the spherical doubly magic lead 
isotope. The two fragments are well defined far before the rupture of 
the neck. The fragments' matter distribution proves that this super 
asymmetric fission channel describes CR. The same topology of the PES 
has been also found in all cluster emitters observed in the light 
actinides \cite{War11}.

The CR valley in $^{258}$No depicted in Fig. \ref{PES} (b) shows the 
same characteristics as in $^{234}$U but it is shifted towards larger 
quadrupole moments. The scission point is located at higher quadrupole 
and octuple deformation making the fission barrier broader. The height 
of the CR fission barrier is reduced to 14.2 MeV. Again, the mass of a 
heavy fragment at scission corresponds to $A=208$. In the super-heavy 
$^{284}$Cn shown in Fig. \ref{PES} (c) the same type of very asymmetric 
fission valley can be found as well. Again, it is shifted towards higher 
quadrupole moments. In this nucleus the traditional symmetric fission 
barrier has a two humped structure and the CR valley starts at the 
minimum located between them instead of the ground state. Moreover, the CR 
fission barrier has substantially changed its shape: it is much lower 
than in previous cases reaching a height of only 1.5 MeV at the saddle 
point. Starting from $Q_{20}=60$ b this fission path drops down below the 
ground state energy. The scission point is located at $Q_{20}=128$ b 
and $Q_{30}=92\;\mathrm{b}^{3/2}$ with an energy 12.6 MeV below ground 
state.

To study the properties of CR in the SH region we have computed PESs 
spanned in the $Q_{20}-Q_{30}$ collective space for all 
aforementioned isotopes and we have found that the CR valley exists in all 
considered nuclei. In Fig. \ref{PATHS} (a) a bunch of CR fission 
paths is shown. Its characteristic pattern presents in the PES
smoothly evolves in going from light actinides to SH nuclei. The only 
sharp modification takes place in the starting point of the paths for $^{268}$Sg and 
heavier nuclei. They start at the second minimum, not at the ground state, 
as it was described above in the case of $^{284}$Cn. In these isotopes 
the CR fission barrier is comprised of two differentiated parts: first a reflection 
symmetric hump followed by a second octupole deformed barrier. In all considered isotopes 
the CR scission configuration contains a spherical heavy mass fragment:
the double magic $^{208}$Pb. This means that in SH elements one may expect decay of the 
same nature as in light actinides.

A similar structure of fission paths in this region was also found with 
the covariant density theory \cite{afanasjevPhysRevC.85.024314} and 
Skyrme energy density functional \cite{staszczakPhysRevC.87.024320} in 
a non-relativistic setup. The asymmetric fragment mass distribution 
predicted here is not in contradiction with symmetric fission barriers 
predicted in macroscopic-microscopic models 
\cite{baranPhysRevC.72.044310,mollerPhysRevC.91.024310,kowalPhysRevC.95.014303}. 
The first saddle point is reflection symmetric and much higher than the 
octupole deformed second one.

The height and shape of fission barriers in super asymmetric fission 
channel are the crucial features for the understanding of CR decay 
lifetimes in heavier nuclei. The CR barriers in the considered nuclei 
are presented in Fig. \ref{PATHS} (b). In the first five isotopes they 
have similar heights of around 25 MeV. The scission point is gradually 
shifted towards higher quadrupole moments, due to the increasing size 
of the cluster in this configuration. Starting from $^{248}$Cf, the 
energy of the scission point gradually decreases. In the mass region 
between 250 and 270 the fission barriers are still very high ($8-20$ 
MeV) and very broad. Tunneling probability across them is too low to 
make it possible to observe the CR channel. The situation changes 
substantially already in $^{274}$Hs. The scission point is below the 
ground state energy and the second, asymmetric, fission barrier is 
lower than 5 MeV, which is less than the energy of the first symmetric 
saddle point. In $^{278}$Ds and heavier nuclei almost the whole 
asymmetric part of the fission path is below the ground state energy. 
In this way, the fission barrier consists basically of the first, 
symmetric, hump which is relatively narrow and easy to tunnel. The 
asymmetric part of fission barrier has little influence on the 
half-lives but it is crucial for the asymmetry of the fragment mass 
distribution.

The reduction of the CR fission barrier height in the region of SH elements came as 
a surprise. However, it can be easy explained on the basis 
of a simple analysis of the Coulomb repulsion energy of two charged spheres at the
scission point. The scission configuration consists of a spherical 
$^{208}$Pb heavy fragment and a lighter cluster created from the remaining nucleons of the
mother nucleus. The separation of the fragments is given by the sum of their radii 
increased by a tip distance $d$. This extra spacing between daughter nuclei comes from the 
neck connecting pre-fragments before scission. 
After the neck rupture, the distance between the position of half-density of the fragments in the post-scission 
configuration vary from 2.5 fm to 4.7 fm. The energy of the scission 
configuration can be estimated as the Coulomb energy of two point charges minus the $Q$ 
value of the decay:
\begin{equation}
E=k\frac{82 (Z-82) e^2}{r_{208}+r_{A-208}+d} - Q\;.
\label{EC}
\end{equation}
Here $Z$ and $A$ are the charge and mass number of the mother nucleus and the
fragment's radius is estimated using the traditional $r_{A_F}=1.2 A_F^{1/3}$ fm expression. 
The results obtained for the considered 
nuclei with an average constant tip distance value $d=3.3$ fm are plotted in Fig. 
\ref{PATHS} (b) with squares. The obtained values perfectly reproduce the 
trend of scission energies and configurations calculated within 
microscopic theory.

\begin{figure}
\includegraphics[height=1\columnwidth, angle=270]{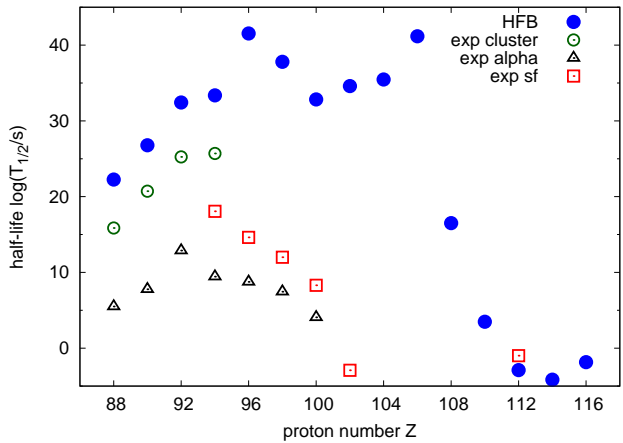}
\caption{CR half-lives of considered isotopes compared with 
experimental data of CR, $\alpha$ emission and spontaneous fission \cite{Audi1674-1137-36-12-001}. \label{T12}}
\end{figure}

The reduction of the CR fission barrier height in the SH region affects the CR 
half-lives plotted in Fig. \ref{T12}. They were calculated using the standard 
WKB approximation. The effective inertia and zero point energy in the
$Q_{20}-Q_{30}$ collective space are calculated using the perturbative cranking approximation for 
pre-scission configuration. After scission, the reduced mass is taken and 
a constant zero point energy is used. Huge values for the
half-lives, longer than $10^{30}$ s are obtained in the $Z=96-106$ region.
However, starting at $^{274}$Hs we observe their substantial reduction. In nuclei with 
$A>280$ we obtain half-lives in the range of what could be measured in 
contemporary SH nuclei experiments. As the calculated half-life for $^{284}$Cn is 
comparable with experimental data for fission we will present a more detailed 
study of this representative isotope.

\begin{figure}
\includegraphics[height=1\columnwidth, angle=270]{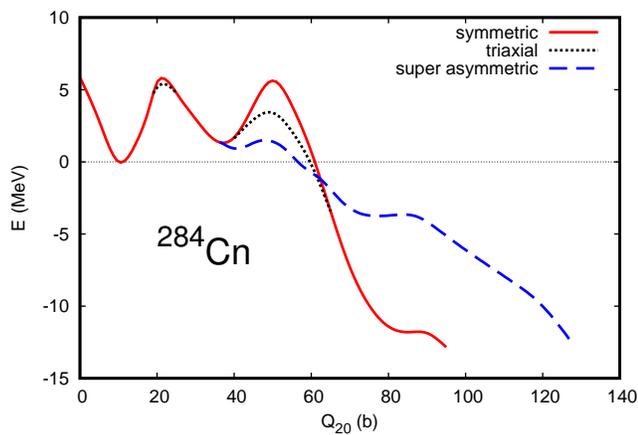}
\caption{Fission barrier of $^{284}$Cn plotted as a function 
of quadrupole moment. \label{bar}}
\end{figure}

In the previous paper \cite{wardaPhysRevC.86.014322} the main decay channel of $^{284}$Cn 
was claimed to be asymmetric fission. Here we have shown that this mode is rather a super asymmetric 
fission mode directly connected with the standard CR observed in the actinides. The other decay 
channels are less favored. Predictions of $\alpha$ radioactivity give 
 4 orders of magnitude longer half-lives. Symmetric fission in 
$^{284}$Cn is also highly suppressed as can be seen in Fig. \ref{PES} (c) and Fig. 
\ref{bar}. The second symmetric barrier has the same width as the 
asymmetric one but it is as high as 5.6 MeV whereas the asymmetric one is 
only 1.5 MeV. Non-axial deformation may reduce the height of the 
fission barrier of SH nuclei but it does not affect the main conclusion. 
First fission barrier is soft against $\gamma$ deformation and its height 
stays practically unchanged (it is diminished only by 0.4 MeV) after including the triaxial degree of freedom. 
The second symmetric barrier is reduced 
as a consequence of triaxiality by around 2.2 MeV only with $\gamma=6^\circ 
$. However, this effect is too weak to make the symmetric 
fission channel the most favorable.
The same conclusion applies to all the isotopes in the region $Z=110-114$ and $N=170-176$ 
centered around $^{284}$Cn that were previously identified as nuclei with asymmetric fission \cite{wardaPhysRevC.86.014322, 
BARAN2015442}. All those nuclei 
should decay in the same mode as $^{284}$Cn. The CR 
fission valley also exists in heavier elements, but decay through this 
channel is suppressed by $\alpha$ emission. From the other hand, in lighter systems the symmetric fission mode is dominant. We 
may conclude that super asymmetric fission, closely related with CR in the 
actinides, can be found in some SH nuclei as the dominant decay mode.

{\it Conclusions.}
We have shown that asymmetric fission in SH nuclei has the 
same nature as CR in light actinides. The dominant decay channel of 
isotopes around $^{284}$Cn is super asymmetric fission with doubly 
magic $^{208}$Pb as the heavy mass fragment. Lighter fragment corresponding
to the SH nuclei discussed here would be $^{70}$Ni, $^{76}$Zn, $^{82}$Ge, and 
$^{86}$Se in the fission of $^{278}$Ds, $^{284}$Cn, $^{290}$Fl, and 
$^{294}$Lv, respectively. It is important to note the existence of the magic numbers 
$Z=28$ and $N=50$ in those light fragments that reinforces the strong influence of 
the magic structure of the heavy fragment. 

The super asymmetric fission mode in SH nuclei discussed here differs from the asymmetric one observed in the Pu-Fm 
region, cf. Fig. \ref{PES} (a). In the actinides, the heavy mass fragment is 
formed by the shell structure of the lighter doubly magic $^{132}$Sn that 
produces a peak in the mass yield at $A_H=140$ \cite{War02}. 
The distinction between asymmetric and super asymmetric fission not only 
concerns the numerical values of most probable fragment masses. Qualitatively 
different shapes of the nucleus can be determined before scission in both 
modes. In CR, the neck is short and narrow whereas in asymmetric fission 
in the actinides it is much longer and thicker \cite{War02}. As neck nucleons 
are shared between fragments at scission 
\cite{wardaPhysRevC.86.024601} the fragment mass yield in 
CR are expected to be much narrower than in asymmetric fission. The 
same conclusion may be deduced from the fact that the CR fission valley is 
very narrow in comparison with the asymmetric one. The variety of 
available fission shapes at the scission line is substantially reduced 
\cite{zdebPhysRevC.95.054608}.

The same mechanism invoked here for spontaneous fission applies 
also to fusion-fission and quasifission observed in this region of SH 
nuclei 
\cite{KozulinPhysRevC.90.054608,kozulinPhysRevC.94.054613,itkis1742-6596-863-1-012043,ITKIS2004136,ITKIS2007150}. 
In all the reactions the possible fragment mass asymmetry is forced by the 
shell structure of $^{208}$Pb.

 It is worth to note that the predicted 
fission fragments produced by CR in SH nuclei lay out of the region of the fission 
products in the actinides i.e. $A=60-180$ with maxima at $A_L\sim100$ and 
$A_H\sim140$. The range of 
possible masses of fragments should be extended in experiments to find signatures of the kind of asymmetric
fission.
Nowadays, experimental techniques do not allow the identification of 
fission fragments in the SH region. The measured energies of the emitted fragments 
show large uncertainties. Only the life-time of these reactions can be 
quite precisely evaluated. We hope that in the nearest future it will 
be possible to determine also masses of the fragments to disentangle the
dominant fission mechanism in these regions. From our 
theoretical analysis we expect that the heavy fragment mass yield should 
be concentrated around $A=208$.

\b
{\it Acknowledgements.}
This work was partly supported by the Polish National Science Centre 
under contracts No. 2016/21/B/ST2/01227 and 2017/24/T/ST2/00396. The 
work of LMR was supported by Spanish grant Nos FPA2015-65929-P MINECO 
and FIS2015-63770-P MINECO.





\bibliographystyle{apsrev}




\end{document}